\documentclass{elsart3p}   

\usepackage{graphics}
\usepackage{graphicx}
\usepackage{dcolumn,bm,fleqn,epic,eepic,times}
\usepackage{amssymb,amsmath,multirow,rotate,color,float}

\begin{document}

\begin{frontmatter}

\title{Networks based on collisions among mobile agents}

\author[1]{Marta C.~Gonz\'alez,}
\author[1,2]{Pedro G.~Lind\corauthref{cor1},}
\corauth[cor1]{Corresponding author: 
               lind@icp.uni-stuttgart.de}
\author[3,4]{Hans J.~Herrmann}
\address[1]{Institute for Computational Physics, 
            Universit\"at Stuttgart, Pfaffenwaldring 27, 
            D-70569 Stuttgart, Germany}
\address[2]{Centro de F\'{\i}sica Te\'orica e Computacional, 
            Av.~Prof.~Gama Pinto 2, 1649-003 Lisbon, Portugal}
\address[3]{Departamento de F\'{\i}sica, Universidade Federal do
            Cear\'a, 60451-970 Fortaleza, Brazil}
\address[4]{IfB, HIF E12, ETH H\"onggerberg, CH-8093 Z\"urich,
             Switzerland}

\begin{abstract}
We investigate in detail a recent model of colliding mobile agents
[Phys.~Rev.~Lett.~96, 088702], used as an alternative approach
to construct evolving networks of interactions formed
by the collisions governed by suitable dynamical rules.
The system of mobile agents evolves towards a quasi-stationary
state which is, apart small fluctuations, well characterized by 
the density of the system and the residence time of the agents.
The residence time defines a collision rate and by varying the collision 
rate, the system percolates at a critical value, with the emergence
of a giant cluster whose critical exponents are the ones of two-dimensional 
percolation.
Further, the degree and clustering coefficient distributions and the average 
path length show that the network associated with such a system presents 
non-trivial features which, depending on the collision rule, enables one
not only to recover the main properties of standard networks, such as
exponential, random and scale-free networks, but also to obtain other 
topological structures.
Namely, we show a specific example where the obtained structure has
topological features which characterize accurately the structure and 
evolution of social networks in different contexts, ranging from networks 
of acquaintances to networks of sexual contacts.
\end{abstract}

\begin{keyword}
Collisions \sep 
Mobile Agents \sep 
Social Contacts \sep 
Complex Networks
\PACS 
89.65.-s \sep 
89.75.Fb \sep 
89.75.Hc \sep 
89.75.Da      
\end{keyword}

\end{frontmatter}


\section{Introduction}
\label{sec:intro}

Till recently, graph theory has been used as the framework
to study the dynamics and the topology of complex systems, where
elementary constituents are represented by nodes and its interactions
mapped to edges\cite{newmanrev,dorogrev,barabasirev,graphsbook}. 
This approach enables the definition of statistical quantities, such
as the clustering coefficient\cite{strogatznat,pre}, the average path
length and the degree distributions\cite{barabasirev}, which characterize  
the network structure and dynamics\cite{dorogrev,graphsbook},
being applied to study physical\cite{krapivsky03},
biological\cite{rogers03,satorras02,chan03} and 
social\cite{davidsen02,socialdistance2,newman02} networks.

There are already many sets of real system data where these
quantities were measured, e.g.~data of virtual\cite{Satorras},
social\cite{NewmanPNAS} and biological networks\cite{Jeong,Oltvai},
and also data of large infrastructure networks\cite{powergrid,airports}.  
However, while for some of these real situations graph theory 
describes the structure and the dynamical evolution of the underlying
system\cite{newmanrev,dorogrev,barabasirev}, there are other situations
where the standard approaches to construct networks do not agree with the 
underlying structure and dynamics\cite{questions,Liljeros,UK}. 
A deeper understanding of these latter situations could be
improved by considering dynamical processes based on local information which
produce such networks~\cite{pre,davidsen02,eisenberg03,prl,epjb}.  

To this end, we proposed recently\cite{prl,epjb} a model where the
nodes (agents) of the networks are mobile, with the connections being
defined through their collisions, i.e.~being a consequence of their
dynamics.   
In other words, the network is build by keeping track of the collisions
between agents, representing the interactions among them. 
Consequently, the network results directly from the time evolution of
the system.
With this approach we are able\cite{prl} to reproduce  
the essential features, namely the dynamical evolution, clustering 
and community structure, observed in empirical networks of social contacts, 
and to explain associated deviations of networks where the contacts are of 
a specific nature, e.g.~sexual\cite{prl,epjb}.
Social networks have attracted significant attention in the
physics community~\cite{davidsen02,socialdistance2,newman02,%
strogatznat01,eames02,freeman}, where, using approaches and techniques
from physical systems, gave further insight to understand
particular social behavior and related phenomena, e.g.~spread of
epidemics~\cite{marta1,marta2}, small-world
effects~\cite{strogatznat,davidsen02}, the mechanisms of social
network growth~\cite{jin01} and cycles of
acquaintances~\cite{pre,bearman04}. 

The main advantages of a model of mobile agents to model structure and
evolution of social acquaintances comes from the fact that 
social interactions should be a result not of an {\it a priori} knowledge 
about the network structure but of local dynamics of the agents from which 
the complex networks emerge.  
Since those social interactions occur among individuals
when they have close affinity or similar social features, it is
understandable to consider networks of social contacts as the result 
of collisions among the agents moving in a continuous space or a 
projection of it whose associated metric is related to a social
distance, defined from the sociologic factors promoting social
contact.

The aim of the present paper is twofold.
On one hand it reviews the model in all its detail and discusses natural
ways to extend it for specific purposes.
On the other hand it shows the features of the mobile agent
model in what concerns quasi-stationary regimes, critical behavior and
topological features.
In Section \ref{sec:model} we describe the model of mobile
agents.
The model attains a quasi-stationary state which is fully 
characterized in Section \ref{sec:qs} and in Section \ref{sec:properties} 
its topological and statistical features are presented.
The system shows a percolation transition which is described in 
Section \ref{sec:percol}.
Application of the agent model to reproduce networks of social and sexual
contacts is presented in Section \ref{sec:addhealth} and finally
Section \ref{sec:conclusions} discusses and concludes the main points
throughout the text.

\section{Model of mobile agents with aging}
\label{sec:model}

In this Section we will show that the model of mobile agents is 
parameterized by two single parameters, the density $\rho$ characterizing its 
composition and the maximal residence time $T_{\ell}$ controlling its 
evolution.
As we will see, other quantities can be derived from these two parameters
for particular purposes, for instance to study the critical behavior of the
system.

The system of mobile agents can be seen as a sort of granular 
gas\cite{refpoeschel}, composed by $N$ particles 
with small diameter $d$, representing agents randomly distributed 
in a two-dimensional system of linear size $L\gg \sqrt{N}d$
and with low density $\rho = N/L^{2} \ll 1/d^2$.
The mobile agents represent individuals moving through the system and 
each pair of agents may collide with each other, given rise to a social 
contact between them. 
This social contact is introduced by setting a link joining the two 
agents after the collision, till its removal when one of the agents leaves 
the system.
Therefore, during the evolution of the system, each agent $i$ is 
characterized by its number $k_i$ of links and by its age 
$A_i$. 
\begin{figure}[t]
\begin{center}
\includegraphics*[width=2.6cm]{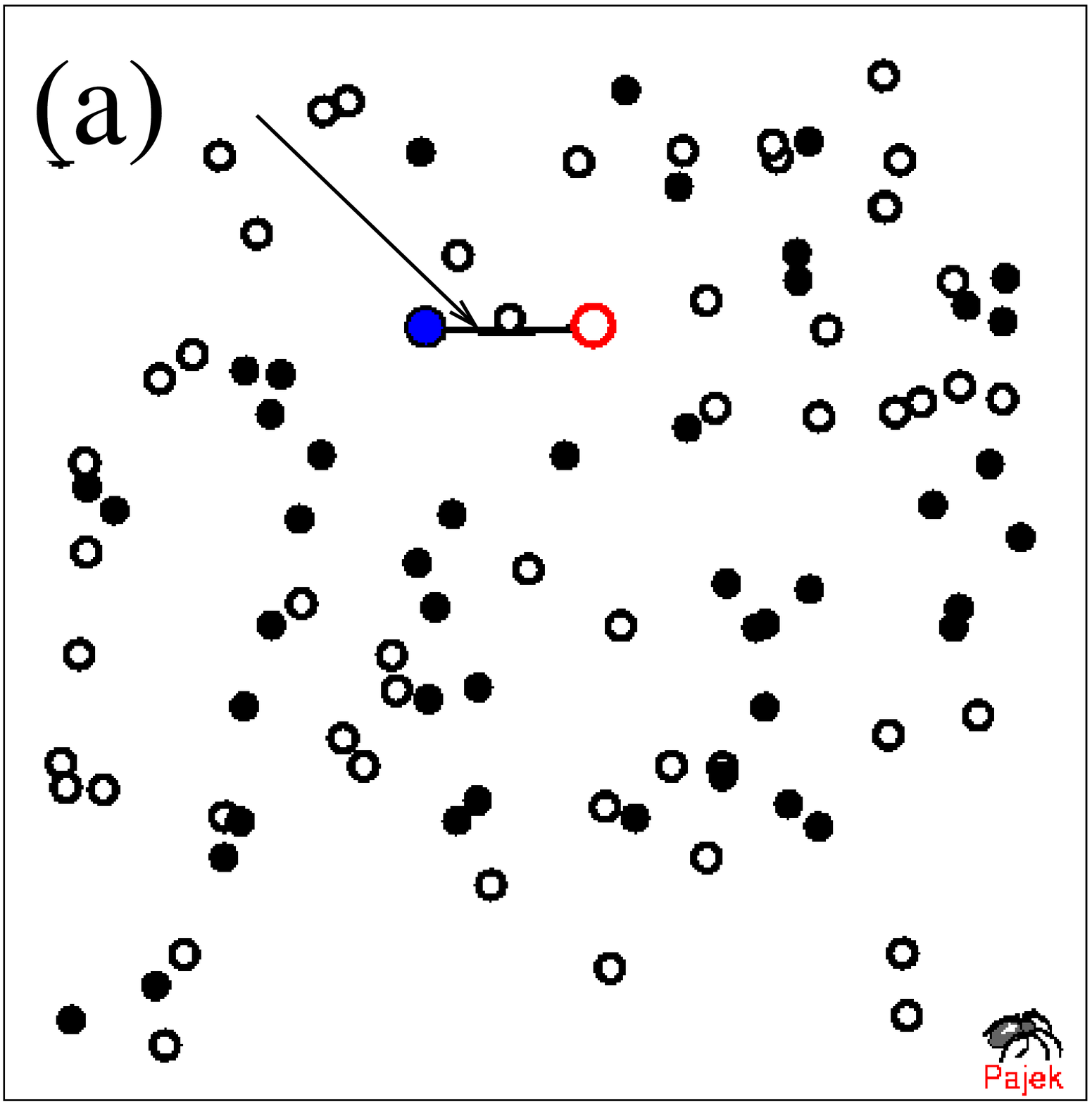}%
\includegraphics*[width=2.6cm]{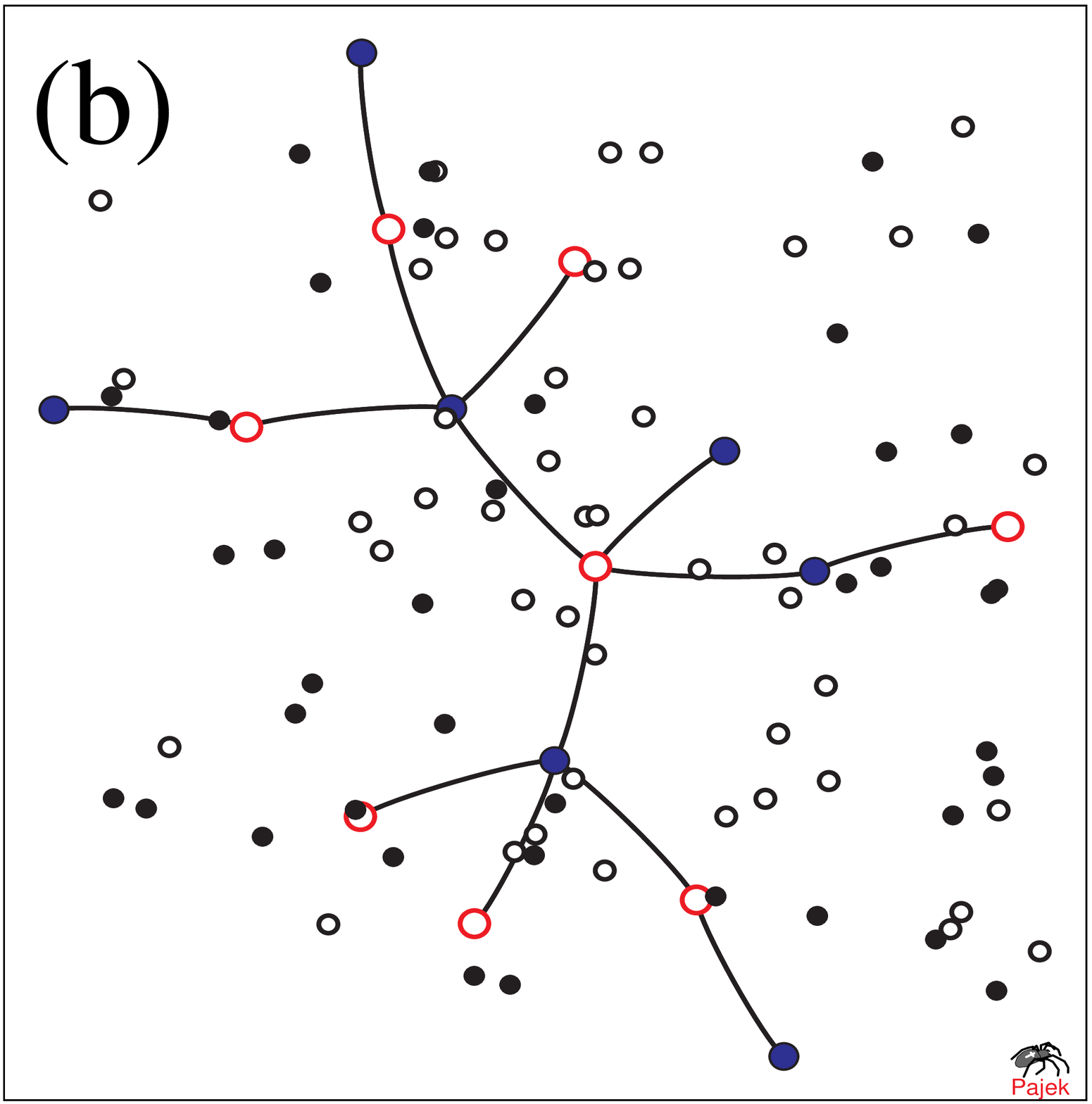}%
\includegraphics*[width=2.6cm]{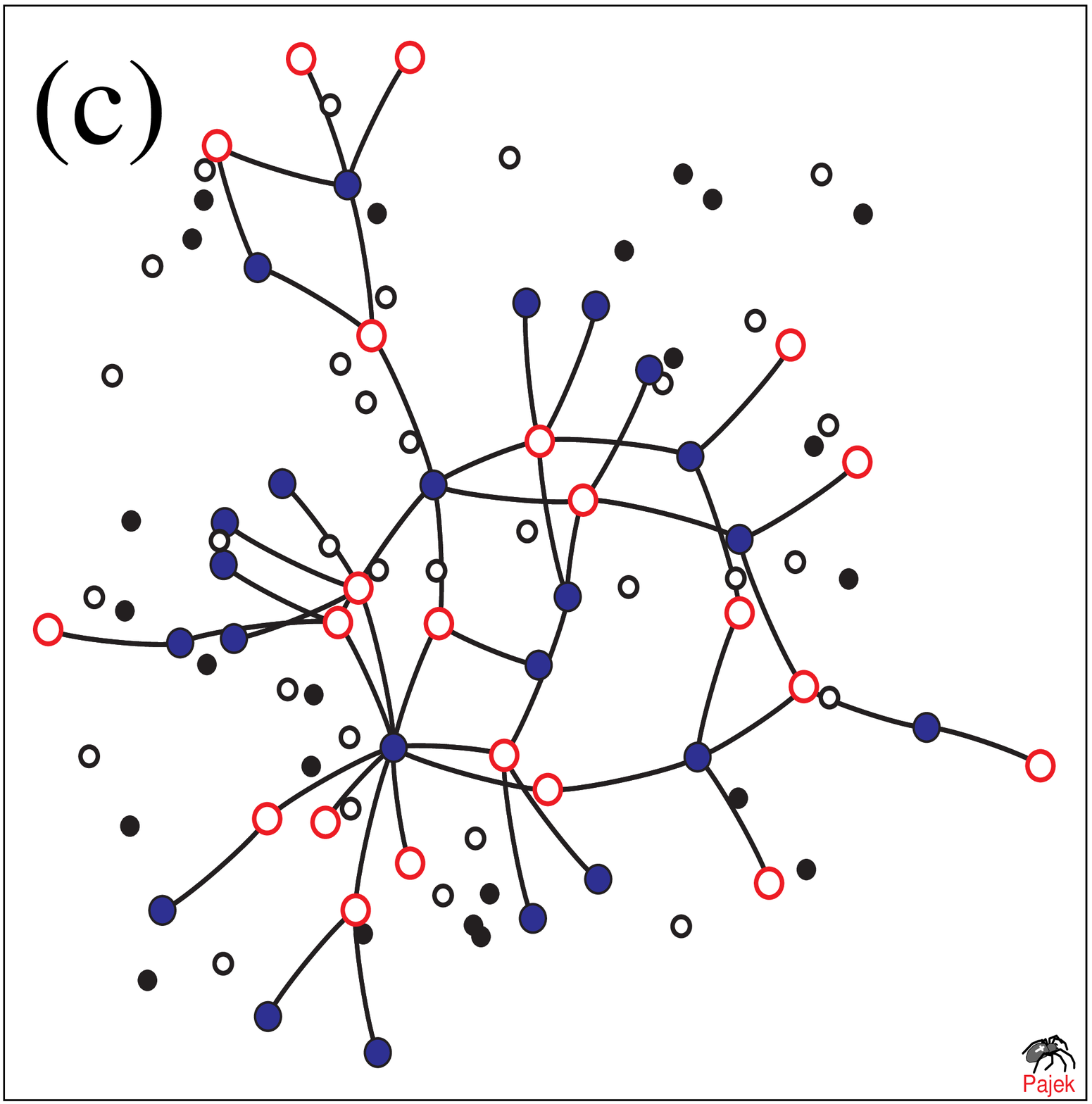}
\end{center}
\caption{\protect
         (Color online)
         Snapshots of the system of mobile agents. Edges between two
         agents indicate that they already collided with each other:
         {\bf (a)} Snapshot after the first collision and
         {\bf (b-c)} two subsequent snapshots within one cluster of
         collisions.
         Filled nodes (blue) and unfilled nodes (red) represent
         two different types of nodes, e.g.~males and females (see
         Sec.~\ref{sec:addhealth} for details).}
\label{fig01}
\end{figure}

We consider periodic boundary conditions and when initialized each
agent has a randomly chosen position and moving direction with
velocity $v_0$.
When for the first time two agents collide, the corresponding
collision is taken as the first contact in the network, as illustrated
in Fig.~\ref{fig01}a (see arrow). 
Letting the agents move and tracking their collisions through time,
more and more connections appear (Figs.~\ref{fig01}b and
\ref{fig01}c).
Figure \ref{fig01} shows the evolution of only one part of the system.
Typically, several networks similar to the ones in Figs.~\ref{fig01}b
and \ref{fig01}c appear during the evolution, each one formed from different
single isolated collisions, as the one illustrated in Fig.~\ref{fig01}a.

The collision process is based on an event-driven algorithm, i.e.~the
simulation progresses by means of a time ordered sequence of collision
events (diffusion) and between collisions each agent follows a ballistic
trajectory\cite{Rapaport} (drift).
According to the algorithm, collisions take place whenever the distance 
between the centers of mass of two agents is equal to their diameter.

Since collisions represent social contacts their dynamical rules should
fulfill some sociological requirements.
Namely, it is known\cite{Laumann} that many social interactions 
occur more commonly between individuals having already a large number
of previous contacts.
For instance, in sexual contacts\cite{epjb} individuals with a larger 
number of partners are more likely to get new partners.
Therefore, we choose a collision rule where the velocity of each agent
may increase with the number $k$ of previous contacts, namely
\begin{equation}
\vec{v}(k_i)=(\bar{v}k_i^{\alpha}+v_0)\vec{\omega} ,
\label{velo}
\end{equation}
where $k_i$ is the total number of social contacts of agent $i$,
$\bar{v}=1\hbox{\ m/s}$ is a constant to assure dimensions of 
velocity, 
$\vec{\omega}=(\vec{e}_x\cos{\theta}+\vec{e}_y\sin{\theta})$ with 
$\theta$ a random angle and $\vec{e_x}$ and $\vec{e_y}$ are unit vectors.

The exponent $\alpha$ in Eq.~(\ref{velo}) controls the velocity update  
after each collision.
For $\alpha=0$ the velocity of each particle is constant in time, and 
consequently the kinetic energy density $E= \tfrac{1}{2}\rho v^2$ of 
the system is constant.
For $\alpha>0$ the velocity increases with degree $k$.
In this range, the value $\alpha=1$ ($\vert\vec{v}\vert\propto k$)
marks a transition between a sub-linear regime ($\alpha<1$) and a 
supra-linear regime ($\alpha>1$) with different degree 
distributions\cite{epjb}.
Throughout the paper we will consider $\alpha=1$ in most of the situations,
showing that it produces the suitable dynamics to reproduce real
networks of social contacts.
It should be noticed that while positive values of $\alpha$ yield 
dynamical laws which fulfill sociological requirements, the 
equation of motion (\ref{velo}) is also able to consider completely
different situations, where $\alpha<0$, i.e.~where the ability to 
acquire new contacts {\it decreases} with the number of previous contacts.
In this manuscript, we will focus in the regimes for $\alpha>0$.

As one may notice, contrary to collision interactions where the velocity 
vector is completely deterministic\cite{ben-Avraham}, here momentum is 
{\it not} conserved.
This is a consequence not only of the increase of $\vec{v}$ but also
of the fact that after one collision the moving direction is randomly 
selected.
The main reason for this random choice is that, as a first approximation 
it is plausible to assume that social contacts do not determine which 
social contact will occur next. 

Concerning the residence time or `age' parameter $A$, during which 
the agents remain in the system, if $A\to\infty$, each agent will eventually 
collide with all the other agents forming a fully connected network.
Whereas, when the average residence time of the agents is finite the system
will reach a non-trivial quasi-stationary 
state\cite{Amaral,dorogovtsev00},
as described in the next Sec.~\ref{sec:qs}.

The aging scheme considered here is simply parameterized by some
threshold in the age of the agents.
More precisely, each agent $i$ is initialized with a certain age
$A_i(0)$ which is a random number uniformly distributed in the
interval $[0,T_{\ell}]$ with $T_{\ell}$ being the maximal age an 
agent may have.
Being updated according to $A_i(t+\Delta t)= A_i(t) + \Delta t$, 
the age eventually reaches $A_{i}(t) = T_{\ell}$,
when the agent $i$ leaves the system, yielding a total residence 
time $T_{\ell}-A_i(0)$.
Computationally the replacement of an old agent by a new one is
carried out simply by removing all the connections of the old
agent and updating its velocity to the initial value $v_0$
with a new random direction. 

When the average residence time is too small, two agents will have no 
time to collide at least once, and consequently no network is formed. 
On the contrary, when $T_{\ell}$ is too large, each agent will 
cross the entire system and a fully connected network appears.
To avoid these two extreme regimes we consider an average residence 
time which is neither very small nor large when compared with the 
characteristic time $\tau$ between collisions.

For that, we define a collision rate, as the fraction between the average
residence time $T_{\ell}-\langle A(0)\rangle=T_{\ell}/2$, 
where $\langle x\rangle$ is the average of $x$ over different
snapshots in the quasi-stationary state, and the 
characteristic time $\tau$ of the mean free path defined as  
\begin{equation}
\tau = \frac{1}{\sqrt{2\pi}\rho d \langle v\rangle} ,
\label{tau}
\end{equation}
where $\langle v\rangle$ is the average velocity of the agents.
With this assumptions our collision rate reads
\begin{equation}
\lambda=\frac{T_{\ell}}{2\tau}= 
        \frac{\langle v\rangle T_{\ell}}{2v_0\tau_0}.
\label{lambda}
\end{equation}
where $\tau_0$ is the characteristic time of the system at the beginning
when all agents have velocity $v_0$.

Before proceeding in characterizing the behavior of such a system it
is important to address three last points to understand the parallel
between the model and real systems.
First, the two-dimensional continuous space where nodes move is {\it not}
the physical space where individuals travel, meet or establish social
acquaintances. 
Instead, it represents a projection of a highly dimensional Euclidean space 
whose metric is related to what is called the social 
distance\cite{socialdistance2,socialdistance1}:
the closer two nodes are, the similar their affinities are (same tastes, same
behavior, etc.) and therefore more probable to establish an 
acquaintance, i.e.~to collide.
It should be stressed that the metric is {\it related} to social distance,
but also incorporates effects of random factors promoting two persons
to meet or establish friendship connections.
We have no rigorous explanation for the fact that a two-dimensional 
projection of such a `large'-dimensional space suffices to reproduce 
empirical data of acquaintances. 
But the fact is that it does, and therefore
for simplicity we will consider a two-dimensional system.
One could also consider higher dimensional systems of mobile agents
but then the `projected' velocity in Eq.(\ref{velo}) would change.

Second, since the system of mobile agents is used to extract a complex
network, the two parameters of the model influence the statistical
features of the network structure. 
Namely, increasing the density 
$\rho$ confines the accessible region of agents thus promoting the 
occurrence of collisions among them which are more
confined in space. In other words, increasing the 
density one increases the clustering coefficient\cite{strogatznat}.
As for $T_{\ell}$, the larger the residence time, the larger the number
of collisions an agent may have. Thus, increasing $T_{\ell}$ increases
the average number $\langle k\rangle$ of connections.

Finally, in a space of affinities what is the meaning of a velocity? 
The velocity is in fact a measure of the accessible region
of a given agent within this social space
which increases, decreases or remains constant
after each collision, depending on the value of $\alpha$.

\section{The quasi-stationary regime}
\label{sec:qs}

In this Section we describe in detail the non-trivial topological and 
dynamical properties of the quasi-stationary (QS) state of the model
described in the previous section.
As we will see, in the QS state the dynamical and topological quantities 
fluctuate around an average value after a transient time of the order 
of $2T_{\ell}$.

\begin{figure}[t]
\begin{center}
\includegraphics*[width=7.8cm]{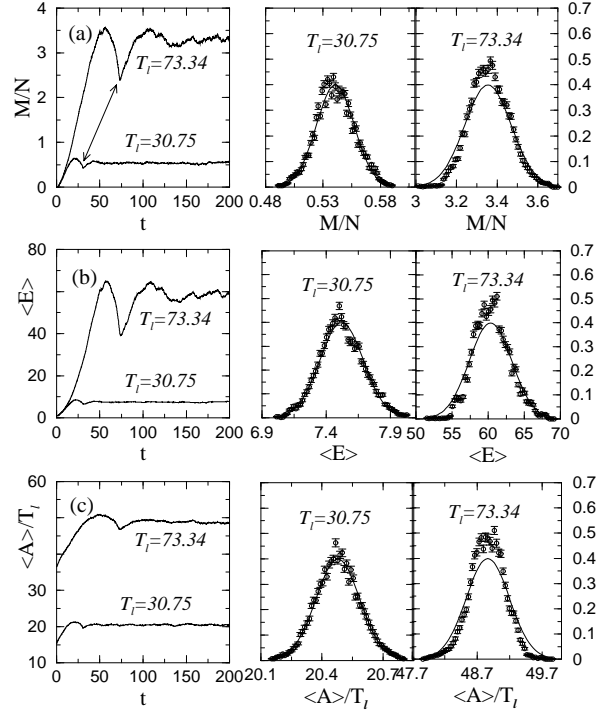}
\end{center}
\caption{\protect 
         Dynamical and topological quantities as function of time
         $t$:
         {\bf (a)} effective coordination $M/N$,
         {\bf (b)} mean energy $\langle E\rangle$,
         {\bf (c)} average age $\langle A\rangle$.
         Two different values of the collision rate are plotted:
         $T_{\ell}=30.75$ and $T_{\ell}=73.34$.
         In (a) arrows indicate $t=T_{\ell}$, the maximal age an agent
         may have. 
         In all cases a quasi-stationary state is attained beyond
         $t\gtrsim 2T_{\ell}$.
         On the right the PDFs of these quantities are shown (circles),
         comparing them with Gaussians having the same central moments
         (solid lines). Here $N=64\times 64$ and $\alpha=1$ 
         (see Eq.~(\ref{velo})).}
\label{fig02}
\end{figure}

Figure \ref{fig02} illustrates the convergence toward the QS state
for two different values of $T_{\ell}$, namely for $T_{\ell}=30.75$
and $T_{\ell}=75.34$.
Here, the convergence is characterized by plotting the effective
coordination given by the fraction $M/N$ between the total number $M$ of
connections and the total number $N$ of nodes (Fig.~\ref{fig02}a), the mean 
energy $\langle E\rangle$ (Fig.~\ref{fig02}b) and the mean age 
$\langle A\rangle$ (Fig.~\ref{fig02}c), as function of time $t$. 
In all three cases, the above quantities increase in the earlier stages 
of the network growth attaining a maximum value around $t\sim T_{\ell}/2$ 
where the agents start dying, resulting in the decrease of their values 
till a minimum at $t = T_{\ell}$.
The maximum at $t\sim T_{\ell}/2$ is due to the fact that the agents
have on average a life-time equal to $T_{\ell}/2$, while the minimum
is due to the extinction of the first `generation' of agents.
Notice that, the maximum occurs at $T_{\ell}/2$ due to the
choice of the initial uniform distribution of the node age which
yields $\langle A(0) \rangle$; in general, for other choice of
age distribution one finds a maximum at $t\sim \langle A(0)\rangle$.
\begin{figure}[t]
\begin{center}
\includegraphics*[width=7.8cm]{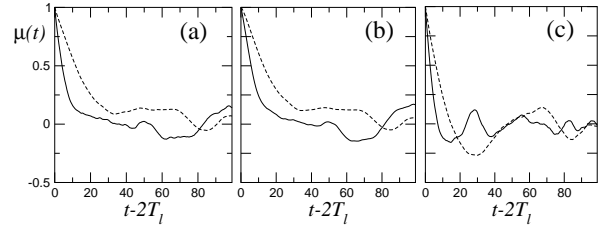}
\end{center}
\caption{\protect 
         Correlation $\mu$ between time step $2T_{\ell}$, beyond which 
         the QS is attained, and a subsequent time-step $t\ge 2T_{\ell}$,
         of the three quantities in Fig.~\ref{fig02}:
         {\bf (a)} effective coordination $M/N$,
         {\bf (b)} mean energy $\langle E\rangle$,
         {\bf (c)} average age $\langle T\rangle$.
         Solid line indicate the result for $T_{\ell}=30.75$, while
         dashed lines indicate the result for $T_{\ell}=73.34$.
         Here $N=64\times 64$ and $\alpha=1$.}
\label{fig03}
\end{figure}

The average values and corresponding fluctuations of these properties
in the QS regime ($t\gtrsim 2T_{\ell}$), depend on $T_{\ell}$, i.e.~on 
$\lambda$ (see Eq.~(\ref{lambda})).
For large values of $\lambda$ (large $T_{\ell}$) the fluctuation is larger
than for small $\lambda$, as can be seen from the probability density
functions (PDF) of each quantity shown on the right of Figs.~\ref{fig02}a-c.
The PDF of the data were computed from the first $2\times 10^4$ data points 
beyond $t\sim 2T_{\ell}$ and are plotted with circles, while a Gaussian with 
the same average and standard deviation is plotted as a solid line
for comparison.
While for $T_{\ell}=30.75$ the distributions are approximate
Gaussian, for $T_{\ell}=73.34$ there is significant deviation from a
Gaussian distribution.

As explained below, between these two values there
is a threshold corresponding to a critical collision rate $\lambda_c$
separating two different behaviors, one where the system is composed
by several small clusters of interconnected nodes and one where all
the nodes belong to a single giant cluster.
Further, from Fig.~\ref{fig03} one concludes that, beyond 
$t\gtrsim 2T_{\ell}$, the correlation length is of the order of $T_{\ell}$.

So, although there is no conservation of momentum, the average velocity 
increasing with time at the beginning, after attaining the QS state 
$\langle v\rangle$ is almost constant.
In the following we assume only values of $\lambda$ 
in the QS state, taking the average velocity in Eq.~(\ref{tau})
as a constant given by the mean of the corresponding PDF distribution.
Further, although all quantities, $M/N$, $\langle E \rangle$ and 
$\langle T\rangle$, characterizing the system in the QS state are 
functions of $T_{\ell}$, we will choose the adimensional parameter 
$T_{\ell}/\tau_0$ instead, which incorporates also the density of the 
system.
\begin{figure}[t]
\begin{center}
\includegraphics*[width=7.8cm]{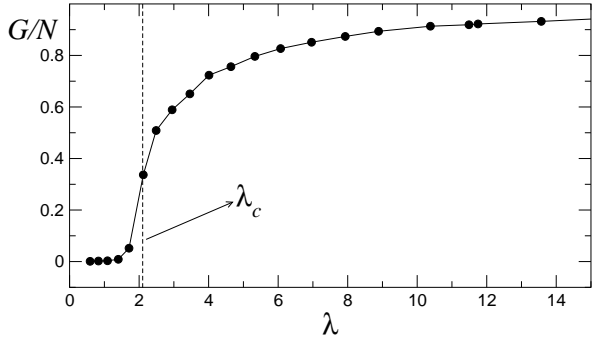}
\end{center}
\caption{\protect 
        Number $G$ of nodes in the largest cluster of the system of
        mobile agents as a function of the collision rate $\lambda$.
        At $\lambda_c\simeq 2.04$ one sees the emergence of a giant
        cluster with $G/N\sim 1$ (see Sec.~\ref{sec:percol}).
        Here $N=128\times 128$ and $\alpha=1$.}
\label{fig04}
\end{figure}

As said above, by varying the collision rate $\lambda$, one finds a 
critical value $\lambda_c$ marking a transition from a state composed by 
several small clusters to a state where a giant cluster emerges 
after attaining the QS state.
In Fig.~\ref{fig04} we plot the fraction $G/N$ between the number 
$G$ of nodes in the largest cluster and the total number $N$ of nodes.
Clearly, beyond $\lambda_c\simeq 2.04$ one sees the emergence of a 
giant cluster.
In the next Section \ref{sec:properties} we describe and discuss
the topological properties for this largest cluster and in Section
\ref{sec:percol} we characterize the percolation transition
occurring at $\lambda_c$.

\section{Properties of the network}
\label{sec:properties}

In this section we will study the degree distribution $P(k)$,
the clustering distributions $C(k)$ characterizing the cliquishness
of the agents neighborhoods, and the average path length $\ell$
in the QS state.

\begin{figure}[t]
\begin{center}
\includegraphics*[width=7.8cm,angle=0]{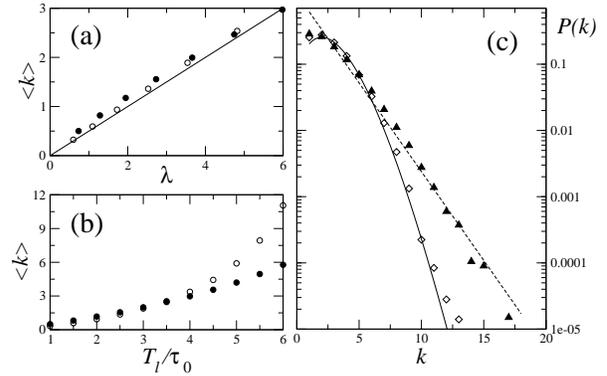}
\end{center}
\caption{\protect 
    Average degree $\langle k\rangle$ as a function of
    {\bf (a)} the collision rate $\lambda$ and of
    {\bf (b)} the maximal age $T_{\ell}/\tau_0$ for
    two values of the density, $\rho=0.02$ (white circles)
    and $\rho=0.2$ (black circles), both with $\alpha=1$.  
    The solid line in (a) indicates $\lambda=2\langle k\rangle$.
    {\bf (c)} Degree distribution of the giant cluster in the QS state 
    for two different velocity updates (see Eq.~(\ref{velo})):
    $\alpha=0$ (triangles) with $T_{\ell}/\tau_0=9.5$, where the
    velocity modulus is always constant and
    $\alpha=1$ (diamonds) with $T_{\ell}/\tau_0=3$, where velocity modulus 
               increases linearly with the number of previous connections.    
    Measuring in each case $\langle k\rangle$ in the system and introducing
    its value in Eqs.~(\ref{poisson}) and (\ref{exponential}), yields the 
    suitable fits for each $\alpha$ value, namely
    a Poisson distribution (solid line) with
    $\langle k \rangle =2.52$ and an exponential distribution
    (dashed line) with $\langle k \rangle =2.62$.
    In all cases, averages over $100$ iterations were taken 
    after attaining the QS state for a system of $N=10^{4}$ agents.}
\label{fig05}
\end{figure}

The average degree $\langle k\rangle$ per agent is a function of both
the collision rate $\lambda$ and rescaled maximal age $T_{\ell}/\tau_0$
as shown for two different values of $\rho$ in Fig.~\ref{fig05}a and 
\ref{fig05}b respectively.
From Fig.~\ref{fig05}a one sees that $\langle k\rangle=\lambda/2$ 
independently on the density $\rho$, while as a function of 
$T_{\ell}/\tau_0$ the average degree increases exponentially with
an exponent depending on the density.
These two plots are important not only to study the topological 
features of the agent model, but also to find the appropriate values 
when modeling real networks as will be explained in Section 
\ref{sec:addhealth}.
\begin{figure}[htb]
\begin{center}
\includegraphics*[width=7.8cm]{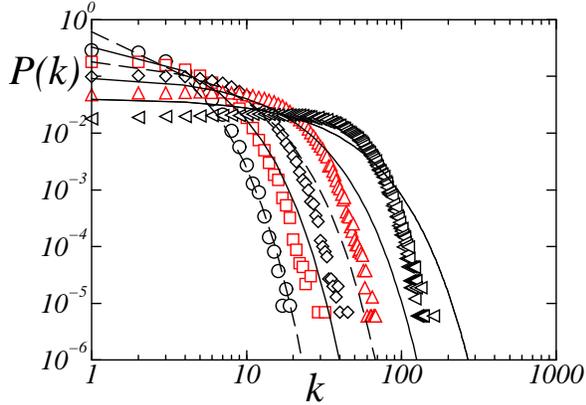}
\end{center}
\caption{\protect          
         {\bf (a)}
         Degree distribution of the giant cluster in the $QS$ state, for 
         several values of $\lambda=5.24, 8.04, 13.04, 23.76$ and $57.36$
         (symbols).
         Lines indicate the corresponding exponential fit
         with Eq.~(\ref{exponential}).
         Here $N=64\times 64$ and $\alpha=1$.
         For other system sizes the results are similar.}
\label{fig06}
\end{figure}
\begin{figure}[htb]
\begin{center}
\includegraphics*[width=7.8cm,angle=0]{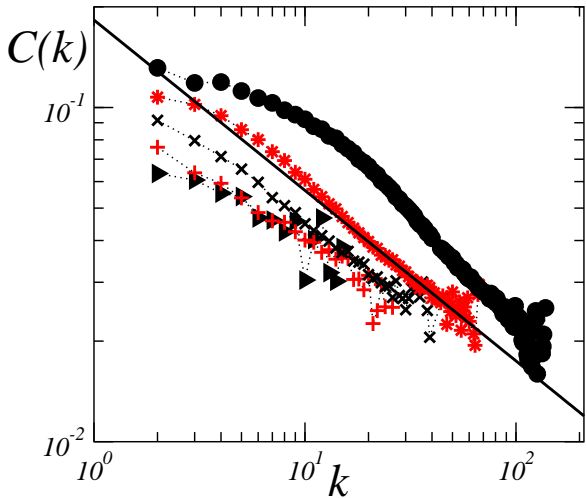}
\end{center}
\caption{\protect
         Clustering coefficient distribution for the same simulations 
         of Fig.~\ref{fig06}, namely for (left to right) 
         $\lambda=5.24, 8.04, 13.04, 23.76$ and $57.36$.
         The solid line is a guide to the eye with slope $1/2$ 
         (see details in text).
         Here $\alpha=1$, $N=128\times128$ and averages over $100$ 
         realizations were taken.}
\label{fig07}
\end{figure}

In Fig.~\ref{fig05}c we compute the degree distribution $P(k)$,
by counting the fraction of nodes having $k$ neighbors.
Two different velocity updates are illustrated here,
namely $\alpha=0$ (triangles) and $\alpha=1$ (diamonds).
Clearly, the degree distribution depends strongly on the collision 
update rule, i.e.~on the value of $\alpha$ in Eq.~(\ref{velo}).

More precisely, for $\alpha=0$, the velocity of each agent is always
constant, the resulting degree distribution being a consequence of the 
effective time\cite{prl} $t_{\hbox{eff}}=3T_{\ell}/8$ to create 
links and of the collision rate, yielding a Poisson distribution
\begin{equation}
P(k)=\frac{\langle k \rangle ^k}{k!}e^{-\langle k \rangle }.
\label{poisson}
\end{equation} 
For this value $\alpha=0$, the degree distribution was 
calculated for a fixed value of $T_{l}/\tau_{0}=9.5$ 
the resulting network has $\langle k \rangle =2.52$, introducing this  
value in eq.\ref{poisson}, we obtain
the solid line in Fig.~\ref{fig05}c, which shows that
the degree distribution of the network is well aproximated 
by a poissonian.
In this way, one can argue that for $\alpha=0$ the system
produces a two-dimensional geometric random graph~\cite{Christensen2} 
in the QS state.

For $\alpha=1$ the velocity in Eq.~(\ref{velo}) increases linearly
with the number of previous collisions. As we will see, this kind of
dynamics reveals to be most suited to reproduce the statistical
features of real social networks (see Sec.~\ref{sec:addhealth}).
In this case, the 
effective residence time is uniformly distributed (see
Sec.~\ref{sec:model}), yielding an exponential velocity distribution 
and consequently an exponential degree distribution 
\begin{equation}
P(k)=\frac{1}{\langle k \rangle -1}e^{-(\frac{k-1}{\langle k \rangle -1})}.
\label{exponential}
\end{equation} 
In Fig.~\ref{fig05} the dashed line indicates the distribution
in Eq.~(\ref{exponential}) with $\langle k \rangle =2.62$, which results
from a value of $T_{l}/\tau_{0}=3.0$ used in the numerical
simulation (triangles).
So, for this value $\alpha=1$ the agent model is capable of producing
a two-dimensional exponential graph in the QS state.

In the following we will consider the value $\alpha=1$ in Eq.~(\ref{velo})
and study how the degree distribution depends on the collision rate 
$\lambda$.
Figure \ref{fig06} shows the degree distribution in the QS state for 
$\lambda=5.24$ (circles), $8.04$ (squares), $13.04$ (diamonds), 
$23.76$ (up triangles) and $57.36$ (left triangles), plotting
with lines the exponential in Eq.~(\ref{exponential}) evaluated with 
the corresponding $\langle k \rangle $. 
As one sees, while for small $\lambda$ the exponential expression fits
well the observed degree distributions, for larger $\langle k \rangle $ 
the numerical results have a lower cutoff than the analytical expression.
Therefore, one concludes that the agent model is able to reproduce
exponential distributions for low values
of the collision rate ($\lambda < 10 $), 
and that other non-trivial distributions 
appear increasing the collision rate.
The latest ones are the ones observed in empirical social networks
(see Sec.~\ref{sec:addhealth}).

Another property of interest to characterize the network
is its clustering coefficient. The local clustering coefficient $C_{i}$ 
of a vertex $i$ with degree $k_{i}$ measures its cliquishness and is 
defined as the quotient between the number $w_i$ of triangles 
(cycles composed by three edges) and the total number of triangles,
\begin{equation}
C_{i}=\frac{2w_{i}}{k_{i}(k_{i}-1)}.
\label{localCC}
\end{equation}
To uncover hierarchical properties of the network, one 
usually\cite{Ravasz} computes the clustering coefficient as a function
of the degree $k$, namely
\begin{equation}
C(k)=\frac{2\langle w(k) \rangle }{k(k-1)},
\label{kCC}
\end{equation}
where $\langle w(k) \rangle $ is the average number of triangles of a 
node with degree $k$. 

\begin{figure}[t]
\begin{center}
\includegraphics*[width=7.8cm]{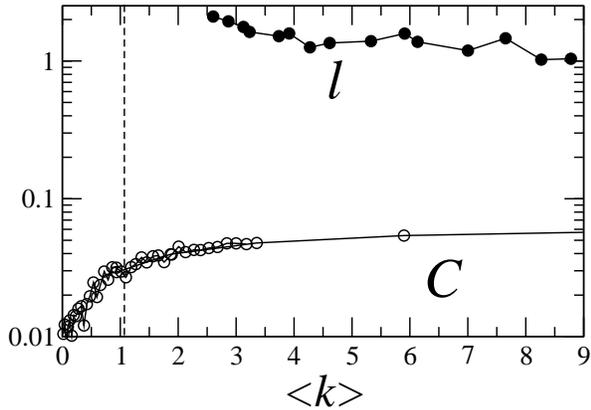}
\end{center}
\caption{\protect 
         Average path length $\ell$ ($\bullet$) and clustering
         coefficients $C$ ($\circ$) as functions of the average
         degree. Here $\rho=0.02$.}
\label{fig08}
\end{figure}

Figure \ref{fig07} shows the mean degree-dependent clustering coefficient
$C(k)$ for the same mobile agents systems of Fig.~\ref{fig06}. 
It is interesting to note that, in contrast to random graphs, which have a 
clustering coefficient independent of $k$, here we observe a dependence of 
the form $\sim k^{-\alpha}$ with $\alpha \in [0.4,0.6]$.
In other words, Fig.~(\ref{fig08}) shows that the clustering coefficient
decreases with the degree, a feature which indicates the existence 
of an underlying hierarchical structure\cite{caldarelli04}.

To end this Section, we study two other topological quantities, namely the 
average shortest path length $\ell$ and the average clustering coefficient 
$C$. As one knows~\cite{strogatznat}, knowing the values
of these topological quantities one is able to ascertain for
small-world effects.

The average clustering coefficient is defined as the average of the 
local clustering coefficients $C_i$ in Eq.~(\ref{localCC}) for all 
$i=1,\dots,N$,
and the average shortest distance $\ell$ is given by the average minimal 
number of edges joining two randomly chosen nodes.

To investigate small-world effects in our system,
we compare the networks of the model of mobile agents with random 
networks having the same number $N$ of nodes and where each pair of nodes is 
connected with a probability $p=2M/(N(N-1))$. 
With this probability one obtains a random network with approximately the 
same effective coordination $M/N$ (same $\langle k\rangle$) as the one
observed in the model of mobile agents.

Figure \ref{fig08} shows these two quantities as a function of the
average degree $\langle k\rangle$.
The clustering coefficient increases with $\langle k \rangle $, and beyond
the emergence of the giant component ($\langle k\rangle>\lambda_c/2=k_c$)
it slowly converges to a value $C_{\infty}\lesssim 0.1$.
The vertical line shows the critical point $k_{c}=1.02$. 
As for the average path length $\ell$ one observes a slow decrease with 
the average degree. Notice that for very low values of $\langle k\rangle$
(not shown) there are no sufficient edges to guarantee a meaningful average 
path length, in particular, below the transition at $k_c$ the system
is divided in several different clusters, which leads to an undefined
$\ell$.
\begin{figure}[t]
\begin{center}
\includegraphics*[width=7.8cm,angle=0]{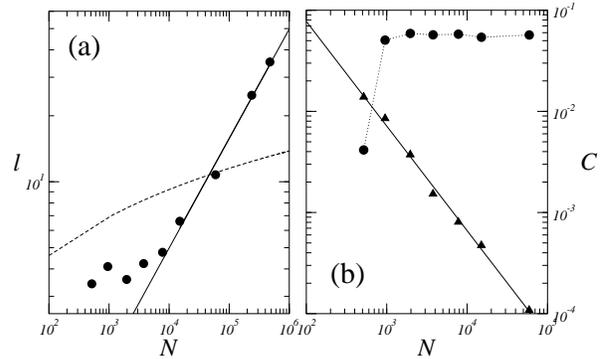}
\end{center}
\caption{\protect
         Size dependence of the networks of mobile agents.
         {\bf (a)} Shortest path length $\ell$ as a function of the system 
         size $N$, for $T_{l}/\tau_{0}=5.02$, 
         compared to $\ln{N}$ (dashed lines). 
         For large network sizes the solid line indicates a fit 
         $\ell = e^{-3}\sqrt{N}$ (see text). 
         {\bf (b)} Average clustering coefficient ($C$) as a function of
         $N$ for the agent model (bullets) compared to the corresponding 
         random graph with the same $\langle k \rangle$ (triangles), having
         a fit (solid line) of $C\sim 1/N$. Here $\rho=0.02$.}
\label{fig09}
\end{figure}

In Fig.~\ref{fig09}a, the average path length is small compared to the 
system size. 
Moreover, as seen in Fig.~\ref{fig09}b the cluster coefficient
in the agent model (circles) is much larger than in the random counterparts
(triangles).
However, the networks generated by the model of mobile agents are not
small-world, since to have the small-world property\cite{barabasirev} it 
is also required that the increase of the shortest path length with the 
system size is not faster than $\ln{N}$.
From Fig.~\ref{fig09}a one sees that this is only true for small
system sizes ($N<10^4$). For larger systems, the fitted numerical results
yield $\ell=e^{-3}\sqrt{N}$.

In Fig.~\ref{fig09}b one sees the behavior of the average cluster coefficient
when the system size is increased.
Interestingly, one clearly sees an independence on $N$, beyond $N>10^3$, where
$C\sim 0.08$. For higher density values $\rho$ this $N$-independent value
of $C$ increases.
This result is quite in contrast to random graphs which scales with $1/N$,
as illustrated in Fig.~\ref{fig09}b with triangles. 

Having characterized the features of the system beyond the transition
at $\lambda_c$, where a giant cluster emerges, in the next Section we 
focus on the properties of this transition.

\section{Critical behavior}
\label{sec:percol}

The emergence of the giant cluster for an exponential network has been 
studied by Molloy and Reed~\cite{Molloy} or equivalently using the generating 
function formalism by Newman ~\cite{Newman}. In both cases
the transition belongs to the universality class of mean field
percolation, in contrast to the agents model which is in the universality class
of two-dimensional percolation, as we will show in this Section. 

The percolation transition is the abrupt change, at a particular
critical value $\lambda_c$, from a state composed by several small
clusters to a state where a giant cluster dominates.
A cluster is defined as a group of connected agents.
Isolated agents are regarded as 
clusters of size unity and any cluster consisting of $s$ connected agents
is called an $s$-cluster. We borrow the notation from Stauffer's book 
on percolation theory~\cite{Stauffer} and define here $n_{s}=N_{s}/N$
as the number of $s$-clusters per agent, where $N_{s}$ is the number
of clusters of size $s$ and $N$ the total number of
agents in the system. 
\begin{figure}[b]
\begin{center}
\includegraphics*[width=7.7cm]{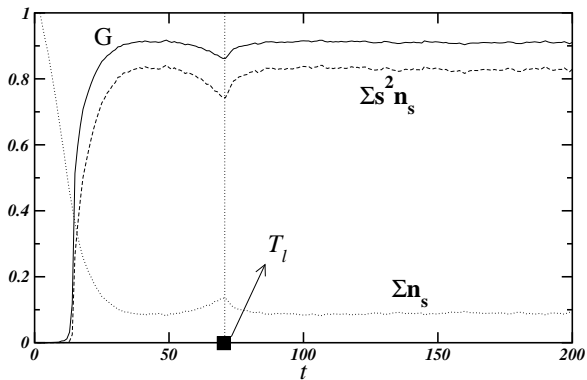}
\end{center}
\caption{\protect 
         Cluster quantities as function of time $t$:
         Number $G$ of agents belonging to the largest cluster (solid line),
         total number of clusters $\Sigma_s n_s$ (dotted line) and
         size-averaged mean cluster size
         $\Sigma_s s^2 n_s$ (dashed line).
         All quantities are normalized by $N$.
         Here $N=128\times 128$ and $\lambda=13.04$.}
\label{fig10}
\end{figure}
\begin{figure}[t]
\begin{center}
\includegraphics*[width=7.8cm,angle=0]{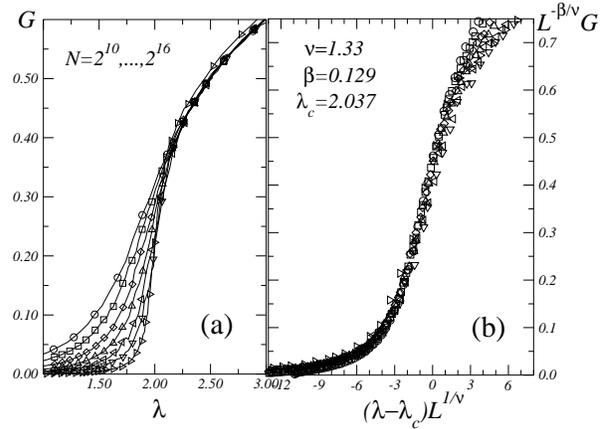}
\end{center}
\caption{\protect
         {\bf (a)} Fractions of sites in the largest cluster $G/N$
         as function of $\lambda$.
         The numerical results where extracted in the $QS$ state for 
         a fixed value of $T_{l}/\tau_{0}$, with $N=32^2,46^2,92^2,128^2$ 
         and $256^2$.
         {\bf (b)} Confirmation of the scaling relation in 
         Eq.~(\ref{eq:coll_big}), for the system sizes 
         $N=2^{10},2^{11},\dots,2^{16}$
         and the values of $\beta$ and $\nu$ reported in 
         Tab.~\ref{table1} at the critical value $\lambda_c=2.04$.}
\label{fig11}
\end{figure}
\begin{figure}[htb]
\begin{center}
\includegraphics*[width=7.8cm,angle=0]{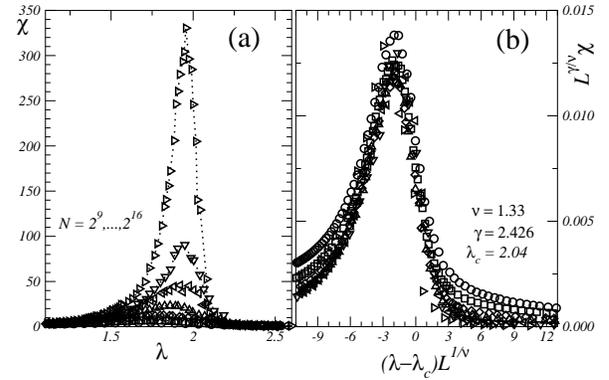}
\end{center}
\caption{\protect
         {\bf (a)} Size-averaged mean cluster size
                   $\sum s^{2}n_{s}$ as a function of
                   $\lambda$ for the same system sizes as in Fig.~\ref{fig11}a.
         {\bf (b)} Confirmation of the scaling relation of 
                   Eq.~(\ref{eq:coll_size}), for the same system sizes 
                   the values of $\gamma$ and $\nu$ reported in 
                   Tab.~\ref{table1} and $\lambda_c=2.04$.}
\label{fig12}
\end{figure}

In Fig.~\ref{fig10} we show the time evolution of the size $G$ of 
the largest cluster (see also Fig.~\ref{fig04}), together with 
the total number $\sum n_{s}$ of clusters and the corresponding 
size-averaged mean cluster size $\sum s^{2}n_{s}$.
At the beginning the size of the giant cluster and the size of the
mean cluster increase till $t=T_{\ell}/2$, corresponding to a complementary
decrease of the number of clusters. Beyond this instant, agents which
have an average residence time of $T_{\ell}/2$, start to leave the
system losing all their connections, and therefore both the size of the
giant cluster and the mean cluster size start to decrease till
$t=T_{\ell}$. Thereafter, all the first generation of nodes has left
the system, and the evolution converges rapidly to the QS state where
all quantities present constant values, apart small fluctuations.
For other values of $T_{\ell}$, i.e.~of the collision rate $\lambda$,
one observes the same behavior of the above quantities, reaching the 
QS state at different values.

The next step is to calculate for a given value of $T_{l}$, the value of
$\lambda$ in the $QS$ state and average at different times the value 
of $G$. The result is plotted in Fig.~\ref{fig11}a for different system
sizes. Simultaneously, the value of $\chi$, given by the size-averaged mean 
cluster size without the largest cluster, is also calculated. The
result is shown in Fig.~\ref{fig12}a. 

In general, according to the standard scaling theory~\cite{Newmanbook}, 
we expect $G$ (Fig.~\ref{fig11}a) and $\chi$ (Fig.~\ref{fig12}a) to follow
\begin{equation}
G=L^{-\beta/\nu} F [(\lambda - \lambda_{c})L^{-1/\nu}]
\label{eq:coll_big}
\end{equation}    
\begin{equation}
\chi=L^{\gamma/\nu} G [(\lambda - \lambda_{c})L^{-1/\nu}].
\label{eq:coll_size}
\end{equation}
where $L$ is the linear size of the system ($N=L^2$).
This is confirmed by the collapses of the curves near the
the critical point, i.e.~$\lambda-\lambda_{c} \sim 0$, 
as illustrated in Figs.~\ref{fig11}b and ~\ref{fig12}b
respectively, for the values of $\nu$, $\beta$, and $\gamma$ 
reported in the central column of Tab.~\ref{table1}, 
with $\lambda_{c}=2.04$.    
\begin{table}[t]
\begin{center}
\begin{tabular}{ccccc} \hline\hline
             &$MF$   & \mbox{mobile agents}   & 
                            \mbox{Percolation (2D)\cite{Stauffer}} \\
$\nu$        & $ 0.5$  &    $1.3 \pm  0.1$     &      $4/3\sim 1.33$ \\
$\gamma$     &  $1$    &    $2.4 \pm  0.1$    &       $43/18\sim 2.39$ \\
$\beta$      &  $1$    &    $0.13 \pm 0.01$  &        $5/36\sim 0.139$ \\
$\sigma$     &  $0.5$  &    $0.40 \pm 0.01$  &        $36/91\sim 0.397$\\
\hline \hline
\end{tabular}
\end{center}
\caption{\protect
         Critical exponents related to the emergence of
         the giant cluster in a random graph model (percolation $MF$),
         for the network of mobile agents presented here, compared
         to the exact results of percolation $2D$.}
\label{table1}
\end{table}

\begin{figure}[htb]
\begin{center}
\includegraphics*[width=7.8cm,angle=0]{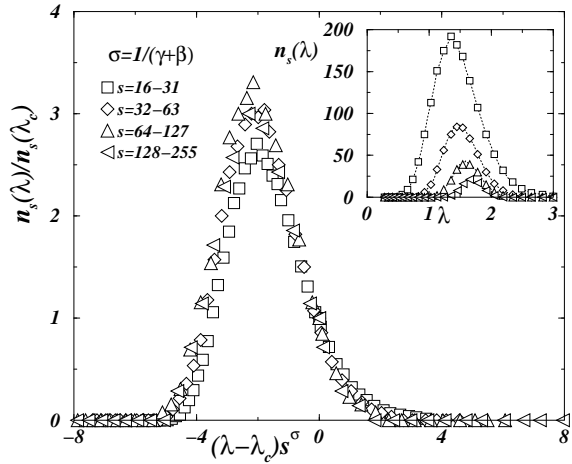}
\end{center}
\caption{\protect
         Collapse of the curves from the inset, plotting 
         $n_{s} / n_{s}(\lambda_{c})$ as a function of 
         $(\lambda - \lambda_{c})s^{\sigma}$ (see text).
         In both cases curves were plotted for cluster sizes
         in the ranges $[16,31]$, $[32-63]$, $[64-127]$, 
         and $[128-255]$.}
\label{fig13}
\end{figure}

In order to test the scaling relation obtained above,
we compute the number $n_s(\lambda)$ of clusters of size $s$
as a function of $\lambda$, as shown in the inset of
Fig.~\ref{fig13}, for $N=2^{16}$ agents.
For the computation, we calculate bins (power-two) of the
cluster size distribution at different values of $\lambda$,
ignoring the first four bins, that is the sizes, $1$, $2-3$,
$4-7$, and $8-15$, since for such small clusters scaling is not
good\cite{Stauffer}. We take the next $4$ bins, i.e.~$s$-values in the ranges
$[16-31]$, $[32-63]$, $[64-127]$ and $[128-255]$ and plot them 
for different $\lambda$.

The main plot in Fig.~\ref{fig13} shows the result after scaling.
Here, for each bin we take the ratio $n_{s}(\lambda)/n_{s}(\lambda_{c})$ 
as a function of $(\lambda - \lambda_{c})s^{\sigma}$ 
using $s$ as the geometric average over the two extremes of the bin ranges
enumerated above.
Performing this scaling we obtain 
\begin{equation} 
\sigma=\frac{1}{\beta+\gamma},
\label{sigma}
\end{equation} 
using the values of $\beta$, $\gamma$ and $\lambda_{c}$ obtained
previously (see Tab.~\ref{table1}). 

Equation (\ref{sigma}), confirms that the scaling relation
of the phase transition belongs to the universality
class of percolation.
It is known that the emergence of a giant cluster
for a random graph depends on $\langle k \rangle $ with its critical 
value at $\langle k \rangle _{c}=1$, and the phase transition belongs
in the same universality class as mean field percolation~\cite{Christensen}, 
whose exponents are given in the first column of Tab.~\ref{table1}.
However, for the agent model one observes the same emergence of the giant 
cluster at $\langle k \rangle _{c}=0.5\lambda_{c}=1.02$ and
the universality class corresponds to $2d$-percolation, as shown in
Tab.~\ref{table1}. 
\begin{figure}[htb]
\begin{center}
\includegraphics*[width=6.5cm,angle=0]{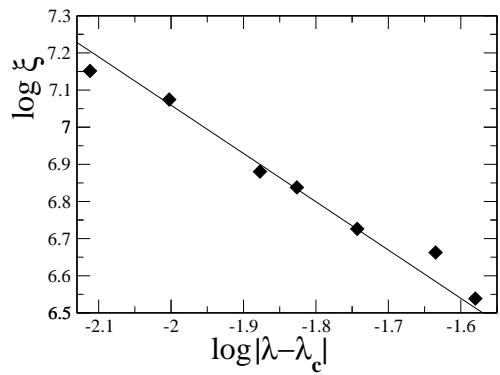}
\end{center}
\caption{\protect
         Correlation length $\xi$ as function of 
         $\lambda-\lambda_c$. Symbols indicate the result of simulations
         performed for different values of $\lambda$ and the solid line
         has a slope of $\nu=-1.3$.}
\label{fig14}
\end{figure}

Before ending this Section, we stress that the correlation exponent 
$\nu$ presented in Tab.~\ref{table1} is obtained from finite 
size scaling. This exponent can be also explicitly calculated from
the linear size of clusters (see \cite{Stauffer}), namely by computing
the correlation length $\xi(\lambda)\sim \vert\lambda-\lambda_c\vert^{-\nu}$.
The result is shown in Fig.~\ref{fig14}, where the solid line has a 
slope of $-1.3$ yielding a correlation exponent in agreement with the
previous results (see Tab.~\ref{table1}). 
Since  the agents move on a two-dimensional plane and have only
a finite life time, they can only establish connections within a
restricted vicinity, and this effect corresponds to a connectivity which
is short range. 
Thus, although the clusters in the agent model are not quenched in time,
the underlying dynamics yields a short range $2d$-percolation.

\section{Real network of social interactions}
\label{sec:addhealth}

In the previous Section we described the main characteristics of the
model of mobile agents in the QS regime, namely an
exponential degree distribution and a transition to percolation for
a certain critical collision rate, beyond which no small-world effects
are observed.
In this Section we analyze a real social network and show that its
topological features are well reproduced by our model. 
In particular, we show that the statistical features observed for the
empirical networks are well reproduced with parameter values beyond
the transition to percolation. 

The empirical data comprehend an extensive study done within the
National Longitudinal Study of Adolescent Health (AddHealth)
\cite{addhealth} at the Carolina Population Center, concerning $84$ 
American schools. The aim is to allow social network researchers
interested in general structural properties of friendship networks to
study the structural and topological properties of social
networks~\cite{bearman04}. 
All data are constructed from the in-school questionnaire from a total
of $90118$ students which responded to it in a survey between 1994
and 1995. The students are separated by the school
they belong to in a total of $84$ schools whose corresponding network
sizes range from $\sim 100$ to $\sim 2000$ students.

\begin{figure}[t]
\begin{center}
\includegraphics*[width=7.8cm]{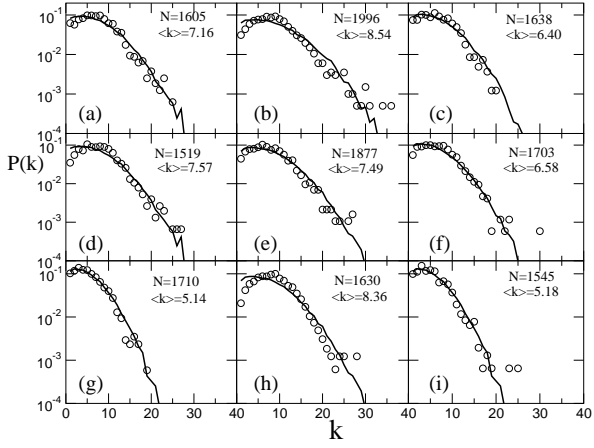}
\end{center}
\caption{\protect 
         Reproducing the distribution of friendship acquaintances in
         empirical social networks~\cite{addhealth} with the agent model. 
         In all cases circles represent the empirical data, while
         solid lines indicate the result of the model of mobile
         agents.}  
\label{fig15}
\end{figure}

Figure \ref{fig15} shows several plots of the degree distributions 
in nine of the schools (circles) compared to the distributions 
obtained with the agent model (solid lines) for the same number of 
agents.
The values for the collision rate in the simulations were taken from 
$\lambda=2\langle k\rangle$ where the average degree is the one found 
in the corresponding school.
As one sees, in all cases the tails of the distributions are well fitted 
by the model. 
\begin{figure}[htb]
\begin{center}
\includegraphics*[width=7.8cm]{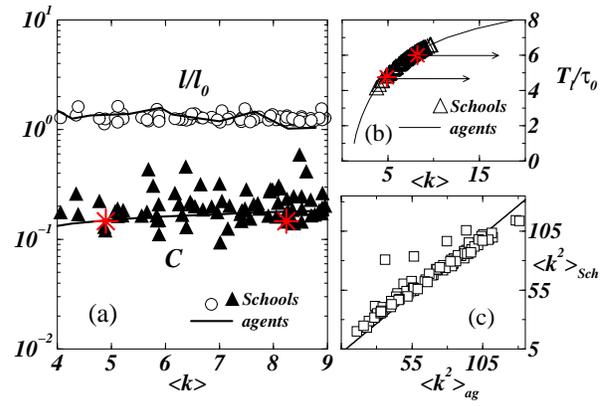}
\end{center}
\caption{\protect 
         (Color online)
         {\bf (a)} Average shortest path length $l$ and clustering
                   coefficient $C$ as functions of the average degree
                   $\langle k \rangle$. Empirical data (symbols) 
                   compared to simulations (solid lines).  
         {\bf (b)} Plot of $T_{l}/\tau_{o}$ as a function of $\langle
                   k \rangle $ for the agents models (solid line). 
                   Stars illustrate two particular schools for 
                   Figs.~\ref{fig15} having 
                   $T_{l}/\tau_{o}=4.75$ (school 1) and $6.0$ (school 2)
                   respectively.
         {\bf (c)} Second moment $\langle k^2\rangle$ for each school 
                   vs.~the second moment of the 
                   corresponding simulation with the agent model
                   (solid line has slope one).
                   In all cases $\rho=0.1$.}
\label{fig16}
\end{figure}

Clearly, the agent model reproduces the degree distribution
observed in schools, by introducing the same average degree found in
real data. The agent model is not only able to reproduce these degree 
distributions, shown in Fig.~\ref{fig15}, but also the corresponding
clustering coefficient and the shortest path length.

Figure \ref{fig16}a shows for each schools (symbols) the average shortest 
path length $\ell$ (circles), and the clustering coefficient (triangles).
Solid lines indicate the results obtained for the agent model using
the same values of $\langle k \rangle $, averaged over
$100$ realizations with $N=2209$ and $\rho=0.1$.
Since $\ell$ depends on the network size, it is divided by the shortest
path length $l_{0}$ of a random graph with the same average degree
and size. 
Clearly, the agent model predicts accurately both the clustering coefficient 
and the shortest path length for the same average degree.

By computing the average degree $\langle k \rangle$ of each
school one is able to obtain the value of $T_{l}/\tau_{0}$
for which the agent model reproduces properly the empirical
data, as illustrated in Fig.~\ref{fig16}b.
Here solid lines indicate the calculated curve from the agent model,
while triangles indicate the values of $T_{l}/\tau_{0}$ chosen to
reproduce the social network of the schools with the resulting
value of $\langle k\rangle$. 
Moreover, the second moment $\langle k^2\rangle_{ag}$ obtained 
with the simulations of the agent model is rescaled by the same
quantity $\langle k^2\rangle_{Sch}$ measured for the empirical school
networks, as shown in Fig.~\ref{fig16}c.

In other words, a strong point of our model is that
since the average degree varies monotonically with
$\lambda$ one can associate a particular value of the
collision rate to each school. 
The collision rate and in particular this parameterization for the
collision rate depends weakly on the network size. Here we
neglect this dependence, as a first approximation.
Using this one-to-one correspondence one is able to compare the
topological quantities between the schools and the model of mobile
agents. 

Notice that,
the Add Health in-school friendship nomination data was constructed
from a questionnaire where each student listed their in-school friends
by order of importance (best friend first) and to each of the friends
they indicate which kinds of contacts they had in the last seven days,
out from five possible choices. According to the number of different
kinds of acquaintances a weight from $1$ to $5$ is attributed to each
social contact. Moreover, since the questionnaires are based on the
information given by one agent, 
the networks are directed.
Here, we each social contact. 
Moreover we assume as a first approximation that social contacts are
unweighted.
Using all the information concerning these empirical
networks~\cite{bearman04} a more detailed study of the structure of these
networks will be presented elsewhere\cite{martanew}.
\begin{figure}[t]
\begin{center}
\includegraphics*[width=7.8cm,angle=0]{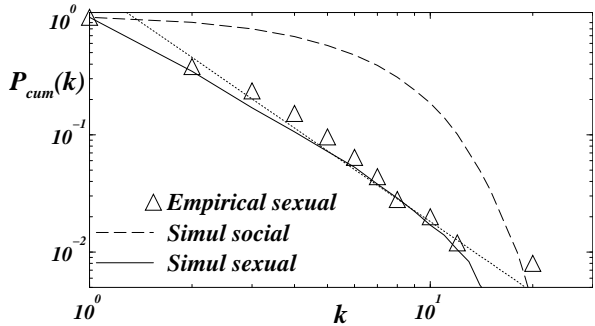}
\end{center}
\caption{\protect
         Cumulative degree distribution of the number $k$ of sexual
         partners in a real empirical network of sexual contacts
         (triangles) with $250$ individuals, compared to the simulation 
         of the agent model
         (solid line), the dotted line is a guide to the eye with
         slope $2$.
         Here $N=4096$, $T_l/\tau_{o}=5.5$ and $\langle k \rangle=7.32$
         and the average size of the resulting sexual network is $220$.}
\label{fig17}
\end{figure}

Finally, we stress that with a slight extension of the model it is possible
to reproduce subsets of social connections, e.g.~sexual contacts, having
particular features different from the entire network of all the 
acquaintances.
For instance, in the case of sexual contacts the degree distribution 
is commonly a power-law~\cite{Liljeros}.
Figure \ref{fig17} shows the degree distribution of a sexual contact network 
(triangles) extracted from a tracing study for HIV tests in Colorado 
Springs\cite{cospring2}.
The dashed line indicates the result obtained for a network of
all social contacts simulated with the agent model,
while the solid line is the degree distribution of 
a subset of contacts from the social network.

The contacts in the subset are chosen by assigning to each agent
an intrinsic property, say a fitness, which enables one to select from all 
the social contacts the ones of particular interest, in this case sexual
contacts.
The fitness is defined by a given number and, when two agents form a link, 
this link is marked as a 'sexual contact' if the sum of agents
fitnesses, $\xi_i + \xi_j$, is greater than a given threshold.
The initial distribution of the fitness is assigned to the agents following 
an exponential distribution and the threshold is $\ln{N}/2$, 
following the scheme of intrinsic fitness proposed in another context by 
Caldarelli et.~al.~\cite{Caldarelli}. 
Interestingly, one is able to extract from the typical distributions of social 
contacts shown in this Section, which are not scale-free, power-law 
distributions in QS state which resemble much the ones observed in real 
networks of sexual contacts (see also Ref.~\cite{epjb}).

\section{Conclusions}
\label{sec:conclusions}

In this paper we introduce an approach to construct networks of
complex interactions from a model of mobile agents. 
The network is constructed by keeping track of the
collision between the mobile agents.
Introducing an aging scheme the network attains a QS
regime, characterized by almost constant topological and dynamical
quantities which fluctuate slightly around an average value.
Typically, the networks constructed with the agent model, while
having small path lengths and large clustering coefficients, 
do not show small-world effects, since the average path length 
does not scale logarithmically with $N$.

The control parameters of the model are the density of the agents
and the maximal residence time the agents may have. With these
two parameters it is possible to define a collision rate which, for
a critical value, marks a percolation transition whose exponents are
the ones of two-dimensional percolation.
While for small collision rates, the degree distributions are 
approximately exponential, for larger values of the collision rate
they strongly deviate from the exponential approximation yielding a 
non-trivial degree distribution.
Surprisingly, these degree distributions are precisely the same as the
ones observed in real social networks of empirical data.
This empirical data has many other informations which were not used in
this study, namely the gender of the agent and the `direction' and
`weight' of the acquaintance. Therefore future investigations on the
context of network analysis of this recent empirical data could be
interesting for sociological purposes\cite{martanew}.

Moreover, we have shown that the agent model is also able to
reproduce networks having only social contacts of particular nature,
e.g.~sexual contacts.
In the particular case of sexual contacts it was found~\cite{Liljeros}
that the distribution of sexual partners has a power-law tail.
The dynamical reasons underlying this discrepancy between the `total'
number of social contacts, with exponential distributions, and the
particular number of sexual contacts, with power-law tails, is not yet
explained. 

\section*{Acknowledgements}

The authors thank M.~Paczuski for useful discussions. 
MCG thanks Deutscher Akademischer Austausch Dienst (DAAD) 
Germany. HJH thanks MPI prize.




\begin{thebibliography}{00}

\bibitem{newmanrev} M.E.J.~Newman,
                    SIAM Rev.~{\bf 45}, 167 (2003). 

\bibitem{dorogrev} S.N.~Dorogovtsev and J.F.F.~Mendes,
                   Adv.~Phys.~{\bf 51}, 1079 (2002).

\bibitem{barabasirev} R.~Albert and A.-L.~Barab\'asi, 
                      Rev.~Mod.~Phys.~{\bf 74}, 47 (2002). 

\bibitem{graphsbook} B.~Bollob\'as,
                     {\it Modern Graph Theory}
                     (Springer, New York, 1998).

\bibitem{strogatznat} D.J.~Watts and S.H.~Strogatz,
                      Nature {\bf 393}, 440 (1998). 

\bibitem{pre} P.G.~Lind, M.C.~Gonz\'alez and H.J.~Herrmann,
              Phys.~Rev.~E {\bf 72}, 056127 (2005);
              cond-mat/0504241.

\bibitem{krapivsky03} P.L.~Krapivsky and S.~Redner,
                      Phys.~Rev.~Lett.~{\bf 90}, 238701 (2003).

\bibitem{rogers03} A.~Rogers,
                   Phys.~Rev.~Lett.~{\bf 90}, 158103 (2003).

\bibitem{satorras02} R.~Pastor-Satorras and A.~Vespignani,
                     Phys.~Rev.~E {\bf 65}, 036104 (2002).

\bibitem{chan03} D.Y.C.~Chan, B.D.~Hughes, A.S.~Leong and W.J.~Reed,
                 Phys.~Rev.~E {\bf 68}, 066124 (2003).

\bibitem{davidsen02} J.~Davidsen, H.~Ebel, and S.~Bornholdt,
                     Phys.~Rev.~Lett.~{\bf 88}, 128701 (2002).

\bibitem{socialdistance2} M.~Bogu\~{n}\'a, R.~Pastor-Satorras, Albert
                          D\'{i}az-Guilera, and Alex Arenas
                          Phys.~Rev.~E. {\bf 70}, 056122 (2004).

\bibitem{newman02} M.E.J.~Newman
                   {Phys.~Rev.~Lett.~}{\bf 89}, 208701 (2002).

\bibitem{Satorras} R.~Pastor-Satorras and A.~Vespignani,
                   {\it  Evolution and Structure of the Internet: A
                   Statistical Physics Approach},  
                   (Princ.~Univ.~Press, Princeton, 1999).

\bibitem{NewmanPNAS} M.E.J.~Newman, 
                     Proc.~Natl.~Acad.~Sci.~{\bf 98},
                     404 (2001).

\bibitem{Jeong} H.~Jeong, B.~Tombor, R.~Albert and A.-L.~Barab\'asi,
                Nature {\bf 407}, 651 (2000).

\bibitem{Oltvai} H.~Jeong, S.P.~Mason, A.-L.~Barab\'asi and
                 Z.N.~Oltvai,
                 Nature {\bf 411}, 41 (2001).

\bibitem{powergrid} D.J.~Watts,
                    {\it Small Worlds: The Dynamics of Networks
                    Between Order and Randomness},
                    (Camb.~Univ.~Press, Cambridge, 2004).

\bibitem{airports} Airport Council International, 
                   ACI Annual Worldwide Airports Traffic Reports 
                   (Airport Counc.~Int., Geneva, 1999).

\bibitem{questions} 
         ``Virtual Round Table on ten leading questions for network research'',
          Eur.~Phys.~J.~B {\bf 38}, 143 (2004).

\bibitem{Liljeros} F.~Liljeros, C.R.~Edling, L.A.N.~Amaral and H.E.~Stanley,
                   Nature {\bf 411}, 907 (2001).

\bibitem{UK} A.~Schneeberger, C.~Mercer, S.A.J.~Gregson,
  N.M.~Ferguson, C.A.~Nyamukapa, R.M.~Anderson, A.M.~Johnson,
  G.P.~Garnett, 
  Sex.~Trans.~Dis.~{\bf 31}, 380 (2004).

\bibitem{eisenberg03} E.~Eisenberg and E.Y.~Levanon,
                   Phys.~Rev.~Lett.~{\bf 91}, 138701 (2003).

\bibitem{prl} M.C.~Gonz\'alez, P.G.~Lind and H.J.~Herrmann
              Phys.~Rev.~Lett.~{\bf 96}, 088702 (2006);
              cond-mat/0602091.

\bibitem{epjb} M.C.~Gonz\'alez, P.G.~Lind and H.J.~Herrmann,
               Eur.~Phys.~J.~B {\bf 49}, 371-376 (2006);
               physics/0508145.

\bibitem{strogatznat01} S.H.~Strogatz,
                        Nature {\bf 410}, 268 (2001).

\bibitem{eames02} K.T.D.~Eames and M.J.~Keeling,
                  Proc.~Nat.~Ac.~Sci.~{\bf 99}, 13330 (2002).

\bibitem{freeman} L.C.~Freeman,
                  {\it The Development of Social Network Analysis}
                  (Vancouver, Canada, 2004).

\bibitem{marta1} M.C.~Gonz\'alez and H.J.~Herrmann, 
                  Physica A {\bf 340} 741 (2004); 
                  cond-mat/0402443.

\bibitem{marta2} M.C.~Gonz\'alez, H.J.~Herrmann and A.D.~Ara\'ujo,
                 Physica A {\bf 356}, 100 (2005); 
                 cond-mat/0502665.

\bibitem{jin01} E.M.~Jin, M.~Girvan, M.E.J.~Newman,
                Phys.~Rev.~E {\bf 64}, 046132 (2001).

\bibitem{bearman04} P.S.~Bearman, J.~Moody and K.~Stovel,
                    Am.J.~of Soc.~{\bf 110}, 44 (2004).

\bibitem{refpoeschel} T.~P\"oschel and S.~Luding,
                      {\it Granular Gases}, 
                      (Lecture Notes in Physics, 564,
                      Springer-Verlag, 2001)

\bibitem{Rapaport} D.C.~Rapaport,
                   {\it The Art of molecular dynamics simulation},
                   (Cambridge University Press, Cambridge, 1995).

\bibitem{Laumann} E.O.~Laumann, J.H.Gagnon, R.T~Michaels,
                    {\it Organization of Sexuality},
                    (University of Chicago Press, 1994).

\bibitem{ben-Avraham} D.~ben-Avraham, E.~Ben-Naim, K.~Lindenberg,
                      A.~Rosas,
                      Phys.~Rev.~E {\bf 68}, 050103 (2003).

\bibitem{Amaral} L.A.N.~Amaral, A.~Scala, M.~Barth\'el\'emy,
                 H.E.~Stanley,
                 Proc.~Natl.~Acad.~Sci.~{\bf 21}, 11149 (2000).

\bibitem{dorogovtsev00} S.N.~Dorogovtsev, J.F.F.~Mendes,
                        Phys.~Rev.~E {\bf 62}, 1842-1845 (2000).

\bibitem{socialdistance1} D.J.~Watts, P.S.~Dodds, M.E.J.~Newman,
                          {Science} {\bf 196}, 1302-1305 (2002).

\bibitem{Christensen2} J.~Dall, M.~Christensen,
                    Phys. Rev. E {\bf 66}, 016121 (2002).

\bibitem{Ravasz} E.~Ravasz, A.-L.~Barab\'asi,
                 Phys.~Rev.~E {\bf 67}, 026112 (2003).

\bibitem{caldarelli04} G.~Caldarelli, R.~Pastor-Satorras and A.~Vespignani,
                       {Eur.~Phys.~J.~B} {\bf 38}, 183-186 (2004).

\bibitem{Molloy} M.~Molloy and B.~Reed, 
                 Comb.~Probab.~Comput.~{\bf 7}, 295 (1998).
       
\bibitem{Newman} D.S.~Callaway, J.E.~Hopcroft, J.M.~Kleinberg, 
                 M.E.J.~Newman and S.H.~Strogatz,
                 Phys.~Rev.~E {\bf 64}, 042902 (2001).

\bibitem{Stauffer} D. Stauffer, 
                   {\em Introduction to Percolation Theory},
                   (Taylor \& Francis, London, 1985).

\bibitem{Newmanbook} M.E.J.Newman and G. T. Barkema
                     {\em Monte Carlo methods in statistical physics},
                     (Clarendon Press, Oxford, 1999).

\bibitem{Christensen} K. Christensen, R. Donangelo, B. Koiler, and K. Sneppen,
                    Phys. Rev. Lett. {\bf 81}, 2380 (1998).

\bibitem{addhealth} This research uses data from Add Health, a program project
designed by J. Richard Udry, Peter S. Bearman, and Kathleen Mullan Harris,
and funded by a grant from the National Institute of Child Health
and Human Developtment (P01-HD31921).        

\bibitem{martanew} M.C.~Gonz\'alez, H.J.~Herrmann, J.~Kert\'esz, T.~Vicsek,
                   in preparation (2006).

\bibitem{cospring2} J.J.~Potterat, L.~Phillips-Plummer, S.Q.~Muth,
                    R.B.~Rothenberg, D.E.~Woodhouse,
                    T.S.~Maldonado-Long, H.P.~Zimmerman, J.B.~Muth,  
                    Sex.~Transm.~Infect.~{\bf 78}, i159 (2002). 

\bibitem{Caldarelli} G.~Caldarelli, A.~Capocci, P.~De Los Rios,
                     M.A.~Mu\~noz, 
                     Phys.~Rev.~Lett.~{\bf 89}, 258702 (2002).

\end{thebibliography}
\end{document}